\documentclass[11pt]{article}

\usepackage[english]{babel}
\usepackage[letterpaper,hmargin=1in,vmargin=1.25in]{geometry}
\usepackage[utf8]{inputenc}
\usepackage{amsmath,amsthm,amssymb,bbm}
\usepackage{graphicx}
\usepackage[colorinlistoftodos]{todonotes}
\usepackage{framed}
\usepackage{stmaryrd}
\usepackage{authblk}
\usepackage{bm}
\usepackage{hyperref}

\newtheorem{thm}{Theorem}
\newtheorem*{thm*}{Theorem}
\newtheorem{prop}[thm]{Proposition}
\newtheorem*{prop*}{Proposition}
\newtheorem{lemma}{Lemma}
\theoremstyle{definition}
\newtheorem{definition}{Definition}

\newtheorem{protocol}{Protocol}

\newtheorem{algo}{Algorithm}
\theoremstyle{remark}
\newtheorem{remark}{Remark}
\theoremstyle{remark*}
\newtheorem*{remark*}{Remark}

\def \bf{\textbf}

\def \mbb{\mathbb}

\def\F{\ensuremath{\mbb{F}}}

\def\C{\ensuremath{\mathcal{C}}}
\def\X{\ensuremath{\mathcal{X}}}
\def\Y{\ensuremath{\mathcal{Y}}}
\def\E{\ensuremath{\mathcal{E}}}
\def\x{\ensuremath{\mathbf{x}}}
\def\y{\ensuremath{\mathbf{y}}}
\def\h{\ensuremath{\mathbf{h}}}

\def\e{\ensuremath{\mathbf{e}}}
\def\w{\ensuremath{\mathbf{w}}}
\def\H{\ensuremath{\mathbf{H}}}

\def\B{\ensuremath{\mathbf{B}}}
\def\lam{\ensuremath{\bm{\lambda}}}
\def\fatmu{\ensuremath{\bm{\mu}}}
\def\d{\ensuremath{\mathbf{\Delta}}}

\title{Perfectly Secure Message Transmission in Two Rounds}

\author[1,2,3]{Gabriele Spini \thanks{Supported by the Algant-Doc doctoral program, \url{www.algant.eu}}}
\author[1]{Gilles Zémor \thanks{Supported by the ``Investments for the Future" Programme IdEx Bordeaux -- CPU (ANR-10-IDEX- 03-02).}}

\affil[1]{Institut de Math\'ematiques de Bordeaux, UMR 5251, Universit\'e de Bordeaux}
\affil[2]{Mathematical Institute, Leiden University}
\affil[3]{CWI Amsterdam}

\date{}

\begin{document}
\maketitle

{\let\thefootnote\relax\footnotetext{
A shorter version of this article, under the same title, \copyright~IACR, is scheduled to appear in the Proceedings of the Theory of Cryptography Conference (TCC) 2016-B.
}}


\begin{abstract}

In the model that has become known as ``Perfectly Secure Message Transmission" (PSMT), a sender Alice is connected to a receiver Bob through $n$ parallel two-way channels. A computationally unbounded adversary Eve controls $t$ of these channels, meaning she can acquire and alter any data that is transmitted over these channels. The sender Alice wishes to communicate a secret message to Bob \emph{privately} and \emph{reliably,} i.e. in such a way that Eve will not get any information about the message while Bob will be able to recover it completely.

In this paper, we focus on protocols that work in two transmission rounds for $n= 2t+1$. We break from previous work by following a conceptually simpler blueprint for achieving a PSMT protocol. We reduce the previously best-known communication complexity, i.e. the number of transmitted bits necessary to communicate a $1$-bit secret, from $O(n^3\log n)$ to $O(n^2\log n)$. Our protocol also answers a question raised by Kurosawa and Suzuki and hitherto left open: their protocol reaches optimal transmission rate for a secret of size $O(n^2 \log n)$ bits, and the authors raised the problem of lowering this threshold. The present solution does this for a secret of $O(n \log n)$ bits.
Additionally, we show how our protocol can be adapted to a Network Coding context.
\end{abstract}


\bf{Keywords}: Perfectly Secure Message Transmission


\section{Introduction}\label{sec:intro}

The problem of Perfectly Secure Message Transmission (PSMT for short) was introduced by Dolev et al. in \cite{Dolev} and involves two parties, a sender Alice and a receiver Bob, who communicate over $n$ parallel channels in the presence of an adversary Eve. Eve is computationally unbounded and controls $t\leq n$ of the channels, meaning that she can read and overwrite any data sent over the channels under her control. The goal of PSMT is to design a protocol that allows Alice to communicate a secret message to Bob \emph{privately} and \emph{reliably,} i.e. in such a way that Eve will not be able to acquire any information on the message, while Bob will always be able to completely recover it. 

Two factors influence whether PSMT is possible and how difficult it is to achieve, namely the number $t$ of channels corrupted and controlled by Eve, and the number $r$ of transmission rounds, where a transmission round is a phase involving only one-way communication (either from Alice to Bob, or from Bob to Alice).

It was shown in Dolev et al.'s original paper \cite{Dolev} that for $r=1$, i.e. when communication is only allowed from Alice to Bob, PSMT is possible if and only if $n\geq 3t+1$. It was also shown in \cite{Dolev} that for $r\geq 2$, i.e. when communication can be performed in two or more rounds, PSMT is possible if and only if $n\geq 2t+1$, although only a very inefficient way to do this was proposed. A number of subsequent efforts were made to improve PSMT protocols, notably in the most difficult case, namely for two rounds and when $n=2t+1$. The following two quantities, called \emph{communication complexity} and \emph{transmission rate}, were introduced and give a good measure of the efficiency of a PSMT protocol. They are defined as follows:

\[
\begin{array}{rl}
\text{Communication complexity}:=& \text{total number of bits transmitted to} \\
&\text{communicate a single-bit secret},
\end{array}
\] 
\[
\text{Transmission rate}:= \frac{\text{total number of bits transmitted}}{\text{bit-size of the secret}}.
\]

Focusing exclusively on the case $n=2t+1$,  Dolev et al.~\cite{Dolev} presented a PSMT protocol for $r=3$ with transmission rate $O\left( n^5 \right)$: for $r=2$ a protocol was presented with non-polynomial rate.

Sayeed and Abu-Amara~\cite{Sayeed} were the first to propose a two-round protocol with a polynomial transmission rate of $O\left( n^3 \right)$. They also achieved communication complexity of $O\left( n^3\log n \right)$. Further work by Agarwal et al.~\cite{Agarwal} improved the transmission rate to $O(n)$ meeting, up to a multiplicative constant, the lower bound of~\cite{Srinathan}. However,
this involved exponential-time algorithms for the participants in the protocol. The current state-of-the art protocol is due to Kurosawa and Suzuki~\cite{KurosawaEuro,KurosawaSuzuki}; it achieves $O(n)$ transmission rate with a polynomial-time effort from the participants. All these protocols do not do better than $O\left( n^3\log n \right)$ for the communication complexity.

We contribute to this topic in the following ways. We present a constructive protocol for which only polynomial-time, straightforward computations are required of the participants, that achieves the improved communication complexity of $O\left(n^2\log n \right)$. In passing, we give an affirmative answer to an open problem of Kurosawa and Suzuki  (at the end of their paper \cite{KurosawaSuzuki}) that asks whether it is possible to achieve the optimal transmission rate $O(n)$  for a secret of size less than $O(n^2 \log n)$ bits. We do this for a secret of $O(n\log n)$ bits. 

Just as importantly, our solution is conceptually significantly simpler than previous protocols. Two-round PSMT involves Bob initiating the protocol by first sending an array of symbols $(x_{ij})$ over the $n$ parallel channels, where the first index $i$  means that symbol $x_{ij}$ is sent over the $i$-th channel. All previous proposals relied on arrays $(x_{ij})$ with a lot of structure, with linear relations between symbols that run both along horizontal (constant $j$) and vertical (constant $i$) lines. In contrast, we work with an array $(x_{ij})$ consisting of completely independent rows $\x^{(j)} = (x_{1j},x_{2j},\ldots ,x_{nj})$ that are simply randomly chosen words of a given Reed-Solomon code. In its simplest, non-optimized, form,  the PSMT protocol we present only involves simple syndrome computations from Alice, and one-time padding the secrets it wishes to transfer with the image of linear forms applied to corrupted versions of the codewords $\x^{(j)}$ it has received from Bob.  

Arguably, the method could find its way into textbooks as a relatively straightforward application of either secret-sharing or wiretap coset-coding techniques. In its optimized form, the protocol retains sufficient simplicity to achieve a transmission rate $5n+o(n)$, compared to the previous record of $6n+o(n)$ of~\cite{Griggio} obtained by painstakingly optimizing the $25n+o(n)$ transmission rate of~\cite{KurosawaSuzuki}.

Finally, we show how our protocol can be adapted to provide security in a more general setting, where Eve can eavesdrop linear combination of the transmitted symbols and inject linear combinations of errors from a set of her choice. This means that thanks to its simpler core structure, our protocol has greater potential of being adapted to more complex scenarios.

\medskip

In the next Section we give an overview of our method and techniques.


\section{Protocol Overview}\label{sec:overview}

The procedure takes as input the number $n=2t+1$ of channels between Alice and Bob and the number $\ell$ of secret messages to be communicated; we assume that the messages lie in a finite field $\F_q$. First, a code $\C$ that will be the basic communication tool is selected; $\C$ is a linear block code of length $n$ over $\F_q$, dimension $t+1$ and minimum distance $t+1$. It furthermore has the property that the knowledge of $t$ symbols of any of its codewords $\mathbf{x}$ leaves $\mathbf{h}\mathbf{x}^T$ completely undetermined, where $\mathbf{h}$ is a vector produced together with $\C$ at the beginning of the protocol. The code $\C$ can be a Reed-Solomon code.

Since we require at most two rounds of communication, Bob starts the procedure; he chooses a certain number of random and independent codewords $\x$, and communicates them by sending the $i$-th symbol of each codeword over the $i$-th channel. This is a first major difference from previous papers, notably \cite{KurosawaSuzuki}, where codewords are communicated in a more complicated ``horizontal-and-vertical" fashion; our construction is thus conceptually simpler and eliminates techniques introduced by early papers \cite{Sayeed} which marked substantial progress at the time but also hindered the development of more efficient protocols when they survived in subsequent work.

As a result of this first round of communication, Alice receives a corrupted version $\mathbf{y}=\mathbf{x}+\mathbf{e}$ for each codeword $\mathbf{x}$ sent by Bob. As in previous PSMT protocols, Alice then proceeds by broadcast, meaning every symbol she physically sends to Bob, she sends $n$ times, once over every channel $i$. In this way privacy is sacrificed, since Eve can read everything Alice sends, but reliability is ensured, since Bob recovers every transmitted symbol by majority decoding.

A secret message consisting of a single symbol $s\in \F_q$ is encoded by Alice as $s + \h\y^T$ for some received vector $\y$. In other words, $s$ is one-time padded with the quantity $\h\y^T$ and this is broadcast to Bob. Notice that at this point, revealing $s + \h\y^T$ to Eve gives her zero information on $s$. This is because she can have intercepted at most $t$ symbols of the codeword $\mathbf{x}$: therefore the element $\mathbf{h}\mathbf{x}^T$ is completely unknown to her by the above property of $\C$ and $\h$, and the mask $\mathbf{h}\mathbf{y}^T= \mathbf{h}\mathbf{x}^T+ \mathbf{h}\mathbf{e}^T$ is unknown to her as well.

Now broadcasting the quantity $s + \h\y^T$ is not enough by itself to convey the secret $s$ to Bob, because Bob also does not have enough information to recover the mask $\h\y^T$. To make the protocol work, Alice needs to give Bob extra information that tells Eve nothing she doesn't already know.

This extra information comes in two parts. The first part is simply the syndrome $\sigma(\y) = \mathbf{H}\mathbf{y}^T$ of $\mathbf{y}$, where $\mathbf{H}$ is a parity-check matrix of $\C$; notice that this data is indeed useless to Eve, who already knows it given that $\mathbf{H}\mathbf{y}^T=\mathbf{H}\mathbf{x}^T+\mathbf{H}\mathbf{e}^T=\mathbf{H}\mathbf{e}^T$ where $\mathbf{e}$ is chosen by herself.

The second part makes use of the fact that during the first phase, Bob has not sent a single codeword $\x$ to Alice, but a batch of codewords $\X$ and Alice has received a set $\Y$ of vectors made up of the corrupted versions $\y = \x +\e$ of the codewords $\x$. Alice will sacrifice a chosen subset of these vectors $\y$ and reveal them completely to Bob and Eve by broadcast. Note that this does not yield any information on the unrevealed vectors $\y$ since Bob has chosen the codewords $\x$ of $\X$ randomly and independently. At this point we apply an idea that originates in \cite{KurosawaSuzuki}: the chosen revealed subset of $\Y$ is called in \cite{KurosawaSuzuki} a \emph{pseudo-basis} of $\Y$. To compute a pseudo-basis of $\Y$, Alice simply computes all syndromes $\sigma(\y)$ for $\y\in \Y$, and chooses a minimal subset of $\Y$ whose syndromes generate linearly all syndromes $\sigma(\y)$ for $\y\in \Y$. A pseudo-basis of $\Y$ could alternatively be called a \emph{syndrome-spanning} subset of $\Y$. Now elementary coding-theory arguments imply that the syndrome function $\sigma$ is injective on the subspace generated by the set of {\em all} errors $\e$ that Eve applies to all Bob's codewords $\x$ (Lemma~\ref{lemma:errweight} and Proposition~\ref{proposition:pseudo-basis}). Therefore a pseudo-basis of $\Y$ gives Bob access to the whole space spanned by Eve's errors and allows him, {\em for any non-revealed $\y=\x +\e$}, to recover the error $\e$ from the syndrome $\sigma(\y)=\sigma(\e)$.

The above protocol is arguably ``the right way'' of exploiting the pseudo-basis idea of Kurosawa and Suzuki, by which we mean it is the simplest way of turning it into a two-round PSMT protocol. We shall present optimized variants that achieve the communication complexity and transmission rate claimed in the Introduction. Our final protocol involves two additional ideas; the first involves a more efficient broadcasting scheme than pure repetition: this idea was also used by Kurosawa and Suzuki. The second idea is new and involves using a decoding algorithm for the code $\C$.

\smallskip

The rest of the paper is organized as follows. In Section~\ref{sec:setting} we recall the coding theory that we need to set up the protocol. In particular, Section~\ref{sec:codes} introduces the code $\C$ and the vector $\mathbf{h}$ with the desired properties. Section~\ref{sec:basis1} introduces Kurosawa and Suzuki's pseudo-basis idea, though we depart somewhat from their original description to fit our syndrome-coding approach to PSMT.

In Section~\ref{sec:protocol1} we describe in a formal way the protocol sketched above, and we compute its communication cost; it will turn out that this construction has a communication complexity of $O\left( n^3 \log n \right)$ and a transmission rate of $O\left( n^2 \right)$.

Section~\ref{sec:improvements} is devoted to improving the efficiency of the protocol; specifically, Section~\ref{sec:genbcast} introduces generalized broadcast, Sections~\ref{sec:basis2} and~\ref{sec:basis3} show how to lower the cost of transmitting the pseudo-basis, while Section~\ref{sec:masks} presents a way to improve the efficiency of the last part of the protocol. A key aspect of this section is that Alice must make extensive use of a \emph{decoding} algorithm for linear codes, a new feature compared to previous work on the topic.

In Section~\ref{sec:protocol2} we implement these improvements and compute the cost of the resulting protocol, reaching a communication complexity of $O\left( n^2 \log n \right)$ and a transfer rate of $5n+o(n)$. We also show in this section that optimal transfer rate is achieved for a secret of $O(n \log n)$ bits.  

Finally, in Section~\ref{sec:linear} we adapt the ideas developed for classical PSMT to a generalized setting with a more powerful eavesdropper. This has applications to achieving security in a Network Coding context.
Section~\ref{sec:conclusion} gives concluding remarks.


\section{Setting and Techniques}\label{sec:setting}


\subsection{Error-Correcting Codes for Communication}\label{sec:codes}

We will use the language of Coding Theory, for background, see e.g. \cite{macwilliams}. Let us briefly recall that when a linear code over the finite field $\F_q$
is defined as $\C=\{\x\in \F_q^n,\, \mathbf{H}\x^T=0\}$, the $r\times n$
matrix $\mathbf{H}$
is called a {\em parity-check matrix} for $\C$ and the mapping 
\begin{eqnarray*}
  \sigma: \F_q^n & \rightarrow & \F_q^r\\
          \x     & \mapsto     & \mathbf{H}\x^T
\end{eqnarray*}
is referred to as the {\em syndrome map}.
 Recall also that a code of parameters (length, dimension, minimum Hamming distance) $[n,k,d]$ is said to be  Maximum Distance Separable or \emph{MDS}, if $d+k=n+1$. Particular instances of MDS codes are Reed-Solomon codes, which exist whenever the field size $q$ is equal to or larger than the length $n$. In a secret-sharing context, Reed-Solomon codes are equivalent to Shamir's secret-sharing scheme~\cite{Shamir}, and they have been used extensively to construct PSMT protocols. We could work from the start with Reed-Solomon codes, equivalently Shamir's scheme, but prefer to use more general MDS codes, not purely for generality's sake, but to stay unencumbered by polynomial evaluations and to highlight that we have no need for anything other than Hamming distance properties.
In section~\ref{sec:protocol2}, we will need our MDS codes to come with a decoding algorithm and will have to invoke Reed-Solomon codes specifically: we will only need to know of the existence of a polynomial-time algorithm though, and will not require knowledge of any specifics.

We will need an MDS code $\C$ that will be used to share randomness,  together with a vector $\h$ such that the value of $\h\x^T$ is completely undetermined for a codeword $\x\in \C$ even when $t$ symbols of $\x$ are known. The linear combination given by $\h$ will then be used to create the masks that hide the secrets.

The following Lemma states the existence of such a pair $(\C,\h)$: it is a slightly non-standard use of Massey's secret sharing scheme~\cite{massey}. It is implicit that we suppose $q> n$, so that MDS codes exist for all dimensions and length up to $n+1$.

\begin{lemma}\label{lemma:privcom}

For any $n$ and any $t< n$ there exists an MDS code $\C$ of parameters $[n,t+1,n-t]$ and a  vector $\mathbf{h}\in \F_q^n$ such that given a random codeword $\x\in \C$, the scalar product $\h\x^T$ is completely undetermined even when $t$ symbols of $\x$ are known.

\begin{proof}
Let $\C'$ be an MDS code of parameters $[n+1,t+1,n-t+1]$; notice that such a code exists for any $n$ and $t\leq n$ \cite{macwilliams}. Let $\C$ be the code obtained from $\C'$ by puncturing at its last coordinate, i.e.
\[
\C:= \left\{ \x\in \F_q^n:\exists \, x\in \F_q \text{ with } (\x,x)\in \C' \right\}
\]
The minimum distance of $\C$ is at most one less than that of $\C'$,
and $\C$ is MDS of parameters $[n,t+1,n-t]$ as requested. Now let
$\H'$ be a parity-check matrix of $\C'$; since $\C'$ has minimum
distance $n-t+1 > 1$, there is at least one row of $\H'$ whose last
symbol is non-zero, i.e. such a row is of the form
\[
(\h, \alpha)\in \F_q^{n+1} \text{ with } \h\in \F_q^n, \, 0\neq \alpha\in \F_q.
\]
We claim that the pair $(\C,\h)$ is of the desired type: indeed, let
$\x$ be a random codeword of $\C$. Then there exists a (unique)
codeword $\x'$ of $\C'$ such that $\x'=(\x,x)$; now $\h\x^T= - \alpha
x$, i.e. the knowledge of $\h\x^T$ is equivalent to the knowledge of $x$, given that $\alpha$ is non-zero.

Now since $\C'$ has dimension $t+1$, for any $t$ known symbols $x_{i_1},\cdots,x_{i_t}$ of $\x$ and any $\tilde{x}\in \F_q$, there exists exactly one $\tilde{\x}'\in \C'$ such that $\tilde{\x}'_{i_j}=x_{i_j}$ for any $j$ and such that $\tilde{\x}'_{n+1}=\tilde{x}$. Hence the claim holds. 

\end{proof}

\end{lemma}

As stated above, this lemma will guarantee the privacy of our protocols; conversely, we can achieve reliable (although not private) communication via the following remark: Alice and Bob can \emph{broadcast} a symbol by sending it over all the channels; since Eve only controls $t< n/2$ of them, the receiver will be able to correct any error introduced by Eve with a simple majority choice. Broadcast thus guarantees reliability by sacrificing privacy.


\subsection{Pseudo-Bases or Syndrome-Spanning Subsets}\label{sec:basis1}

The second fundamental building block of our paper is the notion of \emph{pseudo-basis}, introduced by Kurosawa and Suzuki \cite{KurosawaSuzuki}. The concept stems from the following intuition: assume that Bob communicates a single codeword $\mathbf{x}$ of an MDS code $\C$ to Alice by sending each of its $n$ symbols over the corresponding channel. Eve intercepts $t$ of these symbols, thus $\C$ must have dimension at least $t+1$ if we want to prevent her from learning $\mathbf{x}$; but this means that the minimum distance of $\C$ cannot exceed $n+1-(t+1)=t+1$, which is not enough for Alice to correct an arbitrary pattern of up to $t$ errors that Eve can introduce.

If, however, we repeat the process for several different $\mathbf{x}^{(i)}$, then Alice and Bob have an important advantage: they know that all the errors introduced by Eve \emph{always lie in the same subset of $t$ coordinates}. Kurosawa and Suzuki propose the following strategy to exploit this knowledge: Alice can compute a pseudo-basis (a subset with special properties) of the received vectors; she can then transmit it to Bob, who will use this special structure of the errors to determine their support.

The key is the following simple lemma:

\begin{lemma}\label{lemma:errweight}

Let $\C$ be a linear code of parameters $[n,k,d]_q$, and let $\H$ be a parity-check matrix of $\C$; let $E$ be a linear subspace of vectors of $\F_q^n$  such that the Hamming weight $w_{\texttt{H}}(\e)$ of $\e$ satisfies $w_{\texttt{H}}(\e)<d$ for any 
$\e\in E$.

We then have that the following map is injective:

\[
\begin{array}{rl}
\sigma_{|E}: E &\rightarrow \F_q^{n-k} \\
\e & \mapsto \H\e^T
\end{array}
\]

\begin{proof}

Simply notice that $\ker\left(\sigma_{|E}\right)=\{\mathbf{0}\}$: indeed, $\ker\left(\sigma_{|E}\right) \subseteq \C$; but by assumption all elements of $E$ have weight smaller than $d$, so that $\ker\left(\sigma_{|E}\right)=\{\mathbf{0}\}$.
 
\end{proof}

\end{lemma}

We can now introduce the concept of pseudo-basis; for the rest of this section, we assume that a linear code $\C$ of parameters $[n,k,d]_q$ has been chosen, together with a parity-check matrix $\H$ and associated syndrome map $\sigma$.

\begin{definition}[Pseudo-Basis \cite{KurosawaSuzuki}]\label{definition:pseudo-basis}

Let $\mathcal{Y}$ be a set of vectors of $\F_q^n$; a \emph{pseudo-basis} of $\mathcal{Y}$ is a subset $\mathcal{W} \subseteq\mathcal{Y}$ such that $\sigma(\mathcal{W})$ is a basis of the syndrome subspace $\langle \sigma(\mathcal{Y})\rangle$.

Notice that a pseudo-basis has thus cardinality at most $n-k$, and that it can be computed in time polynomial in $n$.

\end{definition}

The following property formalizes the data that Bob can acquire after he obtains a pseudo-basis of the words received by Alice:

\begin{prop}[\cite{KurosawaSuzuki}]\label{proposition:pseudo-basis}

Let $\X, \E, \Y$ be three subsets:

\[
\begin{array}{rll}
\mathcal{X} := & \left\{ \mathbf{x}^{(1)},\cdots,\mathbf{x}^{(r)}  \right\} & \subseteq \C, \\
\mathcal{E} := & \left\{ \mathbf{e}^{(1)},\cdots,\mathbf{e}^{(r)}
                 \right\} & \subseteq \F_q^n\quad \text{such that } \#\bigcup\left(\texttt{support}\left( \e^{(j)} \right):j=1,\cdots,r\right)< d, \\
\mathcal{Y} := & \left\{ \mathbf{y}^{(1)},\cdots,\mathbf{y}^{(r)}
                 \right\} & \subseteq \F_q^n\quad \text{with } \y^{(j)}=\x^{(j)}+\e^{(j)} \text{ for every } j
\end{array}
\]

Then, given knowledge of $\X$ and a pseudo-basis of $\Y$, we can compute $\mathbf{e}^{(j)}$ from its syndrome $\sigma(\mathbf{e}^{(j)})$, for any $1\leq j\leq r$.

\begin{proof}

The hypothesis on the supports of the elements of $\E$ implies that the subspace $E=\langle\E\rangle$ satisfies the hypothesis of Lemma~\ref{lemma:errweight} and the syndrome function is therefore injective on $\langle\E\rangle$. Given the pseudo-basis $\left\{ \mathbf{y}^{(i)} :i\in I \right\}$, we can decompose any syndrome $\sigma(\e^{(j)})$ as

\begin{align*}
  \sigma(\e^{(j)}) &= \sum_{i\in I} \lambda_i\sigma(\y^{(i)})
                   = \sum_{i\in I} \lambda_i\sigma(\e^{(i)})\\
                   &= \sigma\left(\sum_{i\in I} \lambda_i\e^{(i)}\right)
\end{align*}

which yields $$\e^{(j)}=\sum_{i\in I} \lambda_i\e^{(i)}$$ by injectivity of $\sigma$ on $E$ (cf. Lemma~\ref{lemma:errweight}).

\end{proof}

\end{prop}

\begin{remark}\label{rmk:basis}

Since the syndrome map induces a one-to-one mapping from $E$ to $\sigma(E)$, we also have that $\left\{ \mathbf{y}^{(i)}:i\in I \right\}$ is a pseudo-basis of $\Y$ if and only if $\left\{ \mathbf{e}^{(i)} :i\in I \right\}$ is a basis of $E=\langle\E\rangle$.

\end{remark}

The reader should now have a clear picture of how the pseudo-basis will be used to obtain shared randomness: Bob will select a few codewords $\mathbf{x}^{(1)},\cdots,\mathbf{x}^{(r)}$ in an MDS code of distance at least $t+1$, then communicate them to Alice by sending the $i$-th symbol of each codeword over channel $i$; Alice will be able to compute a pseudo-basis of the received words, a clearly non-expensive computation, then communicate it to Bob. Bob will then be able to determine any error introduced by Eve just from its syndrome as just showed in Proposition~\ref{proposition:pseudo-basis}.

\medskip

The following section gives all the details.


\section{A First Protocol}\label{sec:protocol1}

We now present the complete version of our first communication protocol, following the blueprint of Section~\ref{sec:overview}.

\begin{framed}

\begin{protocol}\label{protocol1}

The protocol allows Alice to communicate $\ell$ secret elements $s^{(1)},\cdots,s^{(\ell)}$ of $\F_q$ to Bob, where $q$ is an arbitrary integer with $q> n$. The protocol takes as input an MDS code $\C$ of parameters $[n,t+1,t+1]_q$ and a vector $\h$ of length $n$ as in Lemma~\ref{lemma:privcom}.

\begin{itemize}

\item[I.] Bob chooses $t+\ell$ uniformly random and independent codewords $\mathbf{x}^{(1)},\cdots,\mathbf{x}^{(t+\ell)}$ of $\C$ and communicates them to Alice by sending the $i$-th symbol of each codeword over the $i$-th channel.

\item[II.] Alice receives the corrupted versions 
$\mathbf{y}^{(1)}=\x^{(1)}+\e^{(1)},\cdots,\mathbf{y}^{(t+\ell)}=\x^{(t+\ell)}+\e^{(t+\ell)}$; she then proceeds with the following actions:

\begin{itemize}

\item[(i)] She computes a pseudo-basis $\left\{\mathbf{y}^{(i)}:i\in I\right\}$ for $I\subset \{1,\cdots,t+\ell\}$ of the received values and broadcasts to Bob $\left(i,\mathbf{y}^{(i)}:i\in I\right)$.

\item[(ii)] She then considers the first $\ell$ words that do \emph{not} belong to the pseudo-basis; to ease the notation, we will re-name them $\mathbf{y}^{(1)},\cdots,\mathbf{y}^{(\ell)}$. For each secret $s^{(j)}$ to be communicated she broadcasts to Bob the following two elements:

\begin{itemize}

\item[-] $\mathbf{H}\left( \mathbf{y}^{(j)} \right)^T$, the syndrome of $\y^{(j)}$;

\item[-] $s^{(j)}+\mathbf{h}\left( \mathbf{y}^{(j)} \right)^T$.

\end{itemize} 

\end{itemize}

\item[III.] 

Proposition~\ref{proposition:pseudo-basis} guarantees that for any $j$, $1\leq j\leq \ell$, Bob can compute the error vector $\e^{(j)}$ and hence reconstruct $\y^{(j)}=\x^{(j)}+\e^{(j)}$ from his knowledge of $\x^{(j)}$. He can therefore open the mask $\mathbf{h}\left( \mathbf{y}^{(j)} \right)^T$ and obtain the secret $s^{(j)}$. 

\end{itemize}

\end{protocol}

\end{framed}

\begin{prop}\label{protocol1:good}

The above protocol allows for private and reliable communication of $\ell$ elements of $\F_q$.

\begin{proof}

As a first remark, notice that since the pseudo-basis has cardinality at most $t$ as remarked in Definition~\ref{definition:pseudo-basis}, Alice has enough words to mask her $\ell$ secret messages, since the total number of words is equal to $t+\ell$. We can now prove that the protocol is private and reliable:

\begin{itemize}

\item \emph{Privacy:} Eve can intercept at most $t$ coordinates of each codeword sent over the channels in the first step; the codewords corresponding to the pseudo-basis are revealed in step~II-(i), but this information is useless since the words are chosen independently and those belonging to the pseudo-basis are no longer used. For any $\mathbf{y}^{(j)}$ that does not belong to the pseudo-basis, the syndrome $\mathbf{H}\left( \mathbf{y}^{(j)} \right)^T$ is also transmitted, but Eve already knows it since 
$\mathbf{H}\left( \mathbf{y}^{(j)} \right)^T=\mathbf{H}\left( \mathbf{x}^{(j)}+\mathbf{e}^{(j)} \right)^T=\mathbf{H}\left( \mathbf{e}^{(j)} \right)^T$, 
where $\mathbf{e}^{(j)}$ denotes the error she introduced herself on $\mathbf{x}^{(j)}$. 

Hence thanks to Lemma~\ref{lemma:privcom}, Eve has no information on any $\mathbf{h}\left( \mathbf{y}^{(j)}\right)^T$, so that privacy holds.

\item \emph{Reliability:} Eve can disrupt the communication only at step~I, since all the following ones only use broadcasts. Proposition~\ref{proposition:pseudo-basis} then ensures that Bob can recover the vectors $\y^{(j)}$ from their syndromes and the corresponding codeword $\x^{(j)}$. From there he can compute and remove the mask $\mathbf{h}\left( \mathbf{y}^{(j)} \right)^T$ without error.

\end{itemize}

\end{proof}

\end{prop}

We now compute the communication complexity and transmission rate of this first protocol, underlining the most expensive parts:

\medskip

\noindent
\emph{Communication complexity:} we can set $\ell:=1$.

\begin{itemize}

\item Step~I requires transmitting $t+1$ codewords over the channels, thus requiring a total of $O\left(n^2\right)$ symbols to be transmitted.

\item

Step~II-(i) requires broadcasting up to $t$ words of $\F_q^n$, thus giving a total of $O\left(n^3\right)$ symbols to be transmitted.

\item Finally, step II-(ii) requires broadcasting a total of $t+1$ symbols (a size-$t$ syndrome and the masked secret), thus giving a total of $O\left(n^2\right)$ elements to be transmitted.

\end{itemize}

Hence since we can assume that $q=O(n)$, we get a total communication complexity of

\[
O\left(n^3 \log n\right)
\]

bits to be transmitted to communicate a single-bit secret.

\medskip

\noindent
\emph{Tranfer rate:} optimal rate is achieved for $\ell=\Omega\left( n\right)$.

\begin{itemize}

\item Step I requires transmitting $t+\ell$ codewords, for a total of $O\left(n^2+n\ell\right)$ symbols.

\item Step II-(i) remains unchanged from the single-bit case, and thus requires transmitting $O\left(n^3\right)$ symbols.

\item Finally, step II-(ii) requires broadcasting a total of $\ell(t+1)$ symbols ($\ell$ size-$t$ syndromes and the masked secrets), thus giving a total of $O\left(n^2\ell\right)$ symbols;

\end{itemize}

To sum up, the overall transmission rate is equal to

\[
\frac{O\left( n^2+n\ell+n^3+n^2\ell \right)}{\ell}=O\left( n^2 \right).
\]

It is immediately seen that the main bottleneck for communication complexity is step II-(i), i.e. the communication of the pseudo-basis, while for transmission rate it is step II-(ii), i.e. the communication of the masked secrets and of the syndromes. We address these issues in the following sections.


\section{Improvements to the Protocol}\label{sec:improvements}

We discuss in this section some key improvements to the protocol; Section~\ref{sec:genbcast} presents the key technique of generalized broadcast, Sections~\ref{sec:basis2} and~\ref{sec:basis3} show a new way to communicate the pseudo-basis (the main bottleneck for communication complexity) and Section~\ref{sec:masks} a new way to communicate the masked secret and the information to open the masks (bottleneck for transmission rate).

\subsection{Generalized Broadcast}\label{sec:genbcast}

Our improvements on the two bottlenecks showed in Section~\ref{sec:protocol1} rely on the fundamental technique of generalized broadcast, which has been highlighted in the paper by Kurosawa and Suzuki \cite{KurosawaSuzuki}.

The intuition is the following: we want to choose a suitable code $\C_{\texttt{BCAST}}$ for perfectly reliable transmission, i.e. we require that if any word $\x\in \C_{\texttt{BCAST}}$ is communicated by sending each symbol $\x_i$ over the $i$-th channel, then $\x$ can always be recovered in spite of the errors introduced by the adversary. In the general situation, since Eve can introduce up to $t$ errors, $\C_{\texttt{BCAST}}$ must have minimum distance $2t+1=n$, and hence dimension 1; for instance, $\C_{\texttt{BCAST}}$ can be a repetition code, yielding the broadcast protocol of Section~\ref{sec:codes}.

Now assume that at a certain point of the protocol, Bob gets to know the position of $m$ channels under Eve's control; then the communication system between the two has been improved: instead of $n$ channels with $t$ errors, we have $n$ channels with $m$ erasures and $t-m$ errors (since Bob can ignore the symbols received on the $m$ channels under Eve's control that he has identified). We can thus expect that reliable communication between Alice and Bob (i.e., broadcast) can be performed at a lower cost by using a code with smaller distance and greater dimension; the following lemma formalizes this intuition.

\begin{lemma}[Generalized Broadcast]\label{lemma:genbcast}

Let $m\leq t$ and let $\C_m$ be an MDS code of parameters $[n,m+1,n-m]_q$; assume that Bob knows the location of $m$ channels controlled by Eve. Then Alice can communicate with perfect reliability $m+1$ symbols $x_1,\cdots,x_{m+1}$ of $\F_q$ to Bob in the following way: she first takes the codeword $\mathbf{c}\in \C_m$ which encodes $(x_1,\cdots,x_{m+1})$, then sends each symbol of $\mathbf{c}$ through the corresponding channel; Eve cannot prevent Bob from completely recovering the message.

We refer to this procedure as \emph{$m$-generalized broadcast}.

\begin{proof}

Notice that $\mathbf{c}$ is well-defined since $\C_m$ has dimension $m+1$. Now since Bob knows the location of $m$ channels that are under Eve's control, he can replace the symbols of $\mathbf{c}$ received via these channels with erasure marks $\bot$, and consider the truncated codeword $\tilde{\mathbf{c}}$ lacking these symbols. Now $\tilde{\mathbf{c}}$ belongs to the punctured code obtained from $\C_m$ by removing $m$ coordinates, which has minimum distance $(n-m)-m \geq 2(t-m)+1$; it can thus correct up to $t-m$ errors, which is exactly the maximum number of errors that Eve can introduce (since she controls at most $t-m$ of the remaining channels). Once he has obtained the shortened codeword $\tilde{\mathbf{c}}$, he can then recover the complete one since $\C_m$ can correct from $m$ erasures, given that it has minimum distance $n-m \geq m$. 
 
\end{proof}  

\end{lemma}

Hence if Alice knows that Bob has identified at least $m$ channels under Eve's control, she can divide the cost of a broadcast by a factor $m$ (since the above method requires to transmit $n$ symbols of $\F_q$ to communicate $m+1$ symbols of $\F_q$).

In the following sections we will make use of Lemma~\ref{lemma:genbcast} to improve the efficiency of the protocol.

\subsection{Improved Transmission of the Pseudo-Basis: a Warm-Up}\label{sec:basis2}

We present here a new method of communicating the pseudo-basis, which is a straightforward implementation of the generalized broadcasting technique.

The key point is the following observation:

\begin{lemma}\label{lemma:basis2}

Let $\mathcal{W}=\left( \mathbf{y}^{(i)}:i\in I \right)$ be a pseudo-basis of the set of received vectors; then if Bob knows $m$ elements of $\mathcal{W}$, he knows at least $m$ channels that have been forged by Eve.

\begin{proof}

By subtracting the original codeword from an element of the pseudo-basis, Bob knows the corresponding error; furthermore, these errors form a basis of the entire error space (Remark~\ref{rmk:basis}). Now if Bob knows $m$ elements of the pseudo-basis, he then knows $m$ of these errors, which necessarily affect at least $m$ coordinates since they are linearly independent. The claim then follows. 

\end{proof}

\end{lemma}

The sub-protocol consisting of the transmission of the pseudo-basis by Alice is simply the following:

\begin{framed}

\begin{protocol}\label{protocol:basis2}

Alice wishes to communicate to Bob a pseudo-basis $\mathcal{W}$ of cardinality $w$.

For any $i=1,\cdots,w$, she then uses $(i-1)$-generalized broadcast to communicate the $i$-th element of the pseudo-basis to Bob.

\end{protocol}

\end{framed}

Lemmas~\ref{lemma:genbcast} and~\ref{lemma:basis2} ensure that this technique is secure; we now compute its cost:

\begin{itemize}

\item Each element of the pseudo-basis is a vector of $\F_q^n$;

\item using $m$-generalized broadcast to communicate $n$ elements of $\F_q$ requires communicating $\left\lceil \frac{n}{m+1} \right\rceil n$ field elements;

\item hence Protocol~\ref{protocol:basis2} requires communicating the following number of elements of $\F_q$:

\[
\sum_{i=1}^w \left\lceil \frac{n}{i} \right\rceil n
= O \left( n^2 \sum_{i=1}^w \frac{1}{i} \right)
= O \left( n^2 \log n \right)
\]

which means that we have reduced to $O\left( n^2 \log^2 n \right)$
the total communication complexity.

\end{itemize}

This complexity is still one logarithmic factor short of our goal; in the next section we show a more advanced technique that allows to bring down the cost to $O\left( n^2 \right)$ field elements.

\subsection{Improved Transmission of the Pseudo-Basis: the Final Version}\label{sec:basis3}

In this section we show a more advanced technique to communicate the pseudo-basis. The key idea is the following: denote by $w$ the size of the pseudo-basis; if Alice can find a received word $\mathbf{y}$ which is subject to an error of weight $cw$ for some constant $c$ and sends it to Bob, then Bob will learn the position of at least $cw$ corrupted channels. Alice will thus be able to use $cw$-generalized broadcast as in Lemma~\ref{lemma:genbcast} to communicate the elements of the pseudo-basis (which amount to $wn$ symbols); since $cw$-generalized broadcast of a symbol has a cost of $O(n/cw)$, the total cost of communicating the pseudo-basis will thus be $(wn)\cdot O(n/cw)=O\left( n^2 \right)$.

We thus devise an algorithm that allows Alice to find a word $\mathbf{y}$ subject to at least $m=\Omega(w)$ errors (for instance, such condition is met if $\mathbf{y}$ is subject to $\Omega(t)$ errors, since $w\leq t$). Notice that such a word $\mathbf{y}$ may not exist among the received words $\left\{ \mathbf{y}^{(i)} \right\}$, therefore we will look for a linear combination of the $\mathbf{y}{(i)}$ with this property. 

As mentioned in Sections~\ref{sec:overview} and \ref{sec:setting}, Alice will make extensive use of a decoding algorithm. Recall that a code of distance $d$ can be uniquely decoded from up to $\left\lfloor (d-1)/2 \right\rfloor$ errors, and that in the case of Reed-Solomon codes, such decoding can be performed in time polynomial in $n$ \cite{macwilliams}; this means that for any Reed-Solomon code $\C$ there exists an algorithm that takes as input a word $\mathbf{y}\in \F_q^n$ and outputs a decomposition $\mathbf{y}=\mathbf{x}+\mathbf{e}$ with $\mathbf{x}\in \C$ and $w_{\texttt{H}}(\mathbf{e})\leq \lfloor (d-1)/2 \rfloor$ (if such a decomposition does not exist, the algorithm outputs an error message $\bot$).

\begin{framed}

\begin{protocol}\label{protocol:basis3}
Alice has received the words $\mathbf{y}^{(1)},\cdots,\mathbf{y}^{(r)}$ and has computed a pseudo-basis $\left\{ \mathbf{y}^{(i)}:i\in I \right\}$ of them; denote by $w$ its cardinality. Alice proceeds with the following actions:

\begin{itemize}

\item she uses Algorithm~\ref{algo:basis3} below to find a ``special word" $\mathbf{y}$, with coefficients $(\mu_i:i\in I)$ such that $\mathbf{y}=\sum_{i\in I}\mu_i \mathbf{y}^{(i)}$. She then communicates to Bob the triplet $\big(I,(\mu_i:i\in I),\mathbf{y}\big)$ by using ordinary broadcast.

\item Finally, she communicates the pseudo-basis of the received values by using $m$-generalized broadcast, where $m:=\min(w,t/3)$, $w$ being the cardinality of the pseudo-basis.

\end{itemize}

\end{protocol}

\end{framed}

Before describing the algorithm formally and proving its validity, we sketch the idea. Alice has computed a pseudo-basis $\{\y^{(i)}: i\in I\}$. For $i\in I$, she applies the decoding algorithm to $\y^{(i)}=\x^{(i)}+\e^{(i)}$. If the decoding algorithm fails, it means that $\y^{(i)}$ is at a large Hamming distance from any codeword, in particular from Bob's codeword $\x^{(i)}$, and the single $\y^{(i)}$ is the required linear combination. If the decoding algorithm succeeds for every $i$, Alice obtains decompositions

\[
\y^{(i)} = \tilde{\x}^{(i)} + \tilde{\e}^{(i)}
\]

where $\tilde{\x}^{(i)}$ is some codeword. Alice must be careful, because she has no guarantee that the codeword $\tilde{\x}^{(i)}$ coincides with Bob's codeword $\x^{(i)}$, and hence that $\tilde{\e}^{(i)}$ coincides with Eve's error vector $\e^{(i)}$. What Alice then does is look for a linear combination $\sum_{i}\mu_i\tilde{\e}^{(i)}$ that has Hamming weight at least $t/3$ and at most $2t/3$. If she is able to find one, then a simple Hamming distance argument guarantees that the corresponding linear combination of Eve's original errors $\sum_{i}\mu_i\e^{(i)}$ also has Hamming weight at least $t/3$. If Alice is unable to find such a linear combination, then she falls back on constructing one that has weight not more than $2t/3$ and at least the cardinality $w$ of the pseudo-basis. This will yield an alternative form of the desired result. We now describe this formally.

\begin{framed}

\begin{algo}\label{algo:basis3}

Alice has a pseudo-basis $\left( \mathbf{y}^{(i)}: i=1,\cdots,w \right)$ (indices have been changed to simplify the notation); the algorithm allows Alice to identify a word $\mathbf{y}$ subject to at least $m:=\min(w,t/3)$ errors introduced by Eve.

In the following steps, whenever we say that the output of the algorithm is a word $\mathbf{y}^{(i)}$, we implicitly assume that the algorithm also outputs the index $i$; more generally, whenever the algorithm outputs a linear combination $\sum_i \mu_i \mathbf{y}^{(i)}$ of the words in the pseudo-basis, we assume that it also outputs the coefficient vector $(\mu_1,\cdots,\mu_w)$ of the linear combination.

\begin{itemize}

\item[1.] Alice uses a unique decoding algorithm to decode the elements of the pseudo-basis; if the algorithm fails for a given word $\mathbf{y}^{(i)}$ (i.e., it doesn't output a codeword having distance at most $t/2$ from $\mathbf{y}^{(i)}$), then Algorithm~\ref{algo:basis3} stops and outputs $\mathbf{y}^{(i)}$.

\item[2.] If the decoding algorithm worked for every $i$, Alice gets a decomposition $\mathbf{y}^{(i)}=\tilde{\mathbf{x}}^{(i)}+\tilde{\mathbf{e}}^{(i)}$ with $\tilde{\mathbf{x}}^{(i)}\in \C$ and $w_H\left( \tilde{\mathbf{e}}^{(i)} \right)\leq t/2$ for every $i$; notice that it is not guaranteed that the $\tilde{\mathbf{x}}^{(i)}$ coincide with the codewords $\mathbf{x}^{(i)}$ originally chosen by Bob.

If any of the $\tilde{\e}^{(i)}$ has weight greater than $t/3$, the algorithm stops and outputs $\mathbf{y}^{(i)}$.

\item[3.] Define $\tilde{\mathbf{f}}^{(1)}:=\tilde{\mathbf{e}}^{(1)}$ and $\tilde{\y}^{(1)}:=\y^{(1)}$. For any $i=2,\cdots,w$, proceed with the following actions:

\begin{itemize}

\item let $\lambda^{(i)}$ be a non-zero element of $\F_q$ such that $\tilde{\mathbf{f}}^{(i-1)}_j+\lambda^{(i)}\tilde{\mathbf{e}}^{(i)}_j\neq 0$ for any coordinate $j\in\{1,2,\ldots ,n\}$ for which $\tilde{\mathbf{f}}^{(i-1)}_j\neq 0$.

\item let $\tilde{\mathbf{f}}^{(i)}:= \tilde{\mathbf{f}}^{(i-1)}+\lambda^{(i)}\tilde{\mathbf{e}}^{(i)}$ and $\tilde{\mathbf{y}}^{(i)}:= \tilde{\mathbf{y}}^{(i-1)}+\lambda^{(i)}\mathbf{y}^{(i)}$;

if $w_{\texttt{H}}\left( \tilde{\mathbf{f}}^{(i)} \right)> t/3$, stop and output $\tilde{\y}^{(i)}$.

\end{itemize}

\item[4.] Output $\tilde{\y}^{(w)}$.

\end{itemize}

\end{algo}

\end{framed}

We can now prove that this algorithm allows Alice to find the desired codeword, which naturally implies that Protocol~\ref{protocol:basis3} indeed allows for reliable communication of the pseudo-basis:

\begin{prop}\label{basis:good}

Algorithm~\ref{algo:basis3} allows Alice to find a word $\mathbf{y}$ subject to an error introduced by Eve of weight at least $m:=\min(w,t/3)$.

\begin{proof}

The following observation is the key point of the algorithm:

\begin{lemma}\label{lemma:weight}

Let $\y=\x+\e=\tilde{\x}+\tilde{\e}$ for $\x,\tilde{\x}\in \C$. Then if $\tilde{\mathbf{e}}$ satisfies $w_{\texttt{H}}(\tilde{\mathbf{e}})\leq 2t/3$, we have that $w_{\texttt{H}}(\mathbf{e})\geq \min \left\{ w_{\texttt{H}}(\tilde{\mathbf{e}}), t/3 \right\}$.

\begin{proof}

The claim is trivial if $\mathbf{e}=\tilde{\mathbf{e}}$; hence assume that $\mathbf{e}\neq \tilde{\mathbf{e}}$. Notice that $\mathbf{e}-\tilde{\mathbf{e}} =\tilde{\mathbf{x}}-\mathbf{x}$; hence since $d_{\texttt{min}}(\C)=t+1$, we have that

\[
t+1 \leq w_{\texttt{H}}\left( \mathbf{e} - \tilde{\mathbf{e}} \right)
\leq w_{\texttt{H}}\left( \mathbf{e} \right) + w_{\texttt{H}}\left( \tilde{\mathbf{e}} \right)
\leq w_{\texttt{H}}\left( \mathbf{e} \right) + \frac{2t}{3}
\]

Hence we have that $w_{\texttt{H}}\left(\mathbf{e} \right) \geq t/3$, so that the claim is proved.
 
\end{proof} 

\end{lemma}

We now analyze the algorithm step-by-step:

\begin{itemize}

\item[1.] if decoding fails for a word $\mathbf{y}^{(i)}$, then it is guaranteed that the error introduced by Eve on it has weight bigger than $t/2> m$ (otherwise, the unique decoding algorithm would succeed since $d_{\texttt{min}}(\C)=t+1$).

\item[2.] since by assumption $w_{\texttt{H}}\left( \tilde{\mathbf{e}}^{(i)} \right)\leq t/2 \leq 2t/3$, if we also have $t/3\leq w_{\texttt{H}}\left( \tilde{\mathbf{e}}^{(i)} \right)$, then thanks to Lemma~\ref{lemma:weight} the output $\mathbf{y}^{(i)}$ is of the desired type.

\item[3.] Since the algorithm did not abort at step 2, all elements $\tilde{\mathbf{e}}^{(i)}$ have weight at most $t/3$. 

First notice that if the algorithm did not produce $\tilde{\mathbf{f}}^{(i-1)}$ as output, then $\tilde{\mathbf{f}}^{(i)}$ is well-defined: indeed, we have that $w_{\texttt{H}}\left( \tilde{\mathbf{f}}^{(i-1)} \right)\leq t/3$; this means that $\lambda^{(i)}$ is well-defined, since it is an element of $\F_q$ that has to be different from 0 and from at most $t/3< n-1$ elements.

Now if the algorithm outputs $\tilde{\mathbf{f}}^{(i)}$, then necessarily $w_{\texttt{H}}\left( \tilde{\mathbf{f}}^{(i-1)} \right)\leq t/3$ (otherwise the algorithm would have stopped before computing $\tilde{\mathbf{f}}^{(i)}$); furthermore, by assumption we have that $w_{\texttt{H}} \left( \tilde{\mathbf{e}}^{(i)} \right)\leq t/3$, so that $w_{\texttt{H}} \left( \tilde{\mathbf{f}}^{(i)} \right)\leq 2t/3$ and we can apply Lemma~\ref{lemma:weight}, so that the output is of the desired type.

\item[4.] Notice that for any $i=1,\cdots,w$, we have that $\tilde{\mathbf{f}}^{(i)}$ has maximal weight among elements of the vector space $\langle \tilde{\mathbf{e}}^{(1)},\cdots,\tilde{\mathbf{e}}^{(i)} \rangle$ (the condition on $\lambda^{(i)}$ ensures that this condition is met at each step). Hence since the elements $\left\{ \tilde{\mathbf{e}}^{(1)},\cdots,\tilde{\mathbf{e}}^{(w)} \right\}$ are linearly independent (because their syndromes are linearly independent, since $(\y^{(1)},\ldots ,\y^{(w)})$ is a pseudo-basis), we have that $w_{\texttt{H}}\left( \tilde{\mathbf{f}}^{(i)} \right)\geq i$ for any $i$.

In particular, we have that $w_{\texttt{H}}\left( \tilde{\mathbf{f}}^{(w)} \right)\geq w$; hence since $w_{\texttt{H}}\left( \tilde{\mathbf{f}}^{(w)} \right)\leq 2t/3$ as remarked above, we have that the output $\tilde{\mathbf{y}}^{(w)}$ is of the desired type.

\end{itemize}

\end{proof}

\end{prop}

\begin{remark}\label{rmk:basiscost}

Protocol~\ref{protocol:basis3} requires Alice to use ordinary broadcast to communicate a single vector of $\F_q^n$ (hence transmitting $n^2$ elements of $\F_q$), then to use $m$-generalized broadcast with $m\geq \min \{w,t/3\}$ to communicate $w\leq t$ vectors of $\F_q^n$ (hence transmitting at most $3n^2$ elements of $\F_q$). We thus get a total of at most $4n^2$ elements of $\F_q$ to be transmitted.

Furthermore, Algorithm~\ref{algo:basis3} has running time polynomial in $n$, as long as the code $\C$ has a unique-decoding algorithm of polynomial running time as well. As already remarked, such algorithms exist for instance for Reed-Solomon codes.

\end{remark}

We study the second bottleneck of the original protocol in the next section.


\subsection{The Improved Communication of the Masked Secrets}\label{sec:masks}

We present in this section the second key improvement to the protocol: after the pseudo-basis is communicated, we devise a way to lower the cost of transmitting to Bob the masked secrets and the information to open the masks. We aim at a cost linear in the number $\ell$ of secrets to be transmitted (while it was quadratic in Protocol~\ref{protocol1}). As in Section~\ref{sec:basis3}, Alice makes use of a unique decoding algorithm.

\begin{framed}

\begin{protocol}\label{protocol:masks}
The protocol is performed once the pseudo-basis has been communicated to Bob; we thus assume that Bob knows the global support $\mathcal{S}:=\cup_i \texttt{support}\left( \e^{(i)} \right)$ of the errors affecting the elements $\mathbf{y}^{(i)}$ (cf. Remark~\ref{rmk:basis}). We assume that Alice wishes to communicate $\ell$ secret elements $s^{(1)},\cdots,s^{(\ell)}$ of $\F_q$ to Bob, and that $\ell$ codewords $\mathbf{x}^{(1)},\cdots,\mathbf{x}^{(\ell)}$ of $\C$ have been sent by Bob to Alice (who has received $\mathbf{y}^{(1)},\cdots,\mathbf{y}^{(\ell)}$) and have not been disclosed in other phases.

\begin{itemize}

\item Alice uses a unique decoding algorithm to decode $\mathbf{y}^{(i)}$, so that for every $i$ she obtains (if decoding was successful) a decomposition $\mathbf{y}^{(i)}=\tilde{\mathbf{x}}^{(i)}+\tilde{\mathbf{e}}^{(i)}$ with $\tilde{\mathbf{x}}^{(i)}\in \C$ and $w_{\texttt{H}}\left( \tilde{\mathbf{e}}^{(i)} \right)\leq t/2$.

For every $i=1,\cdots,\ell$ she then communicates the following elements to Bob:

\begin{itemize}

\item the syndrome $\mathbf{H}\left( \mathbf{y}^{(i)} \right)^T$ via $t/2$-generalized broadcast;

\item the elements $z_1^{(i)},z_2^{(i)}$ of $\F_q$ by ordinary broadcast, where

\[
\begin{array}{cl}
z_1^{(i)} :=& \quad \, s^{(i)}+\mathbf{h}\left( \mathbf{y}^{(i)}\right)^T \\
z_2^{(i)} :=& \left\{
\begin{array}{ll}
s^{(i)}+\mathbf{h}\left( \tilde{\mathbf{x}}^{(i)}\right)^T & \text{ if decoding succeeded}, \\
0 & \text{ otherwise.}
\end{array}
\right.
\end{array}
\]

\end{itemize}

\item Bob can then obtain each secret $s^{(i)}$ in a different way depending on the size of the global support $\mathcal{S}$ of the errors:

\begin{itemize}

\item if $|\mathcal{S}| \geq t/2$, he uses the knowledge of the syndrome of $\mathbf{y}^{(i)}$ and of the support of the error to compute $\mathbf{y}^{(i)}$, so that he can compute $z_1^{(i)} - \mathbf{h}\left( \mathbf{y}^{(i)} \right)^T$ as well.

\item if $|\mathcal{S}| < t/2$, he ignores the syndrome that has been communicated to him, and computes $z_2^{(i)} - \mathbf{h}\left( \mathbf{x}^{(i)} \right)^T$.

\end{itemize}
\end{itemize}
\end{protocol}
\end{framed}

We now prove that this protocol works and is secure:

\begin{prop}\label{masks:good}

The above protocol allows for private and reliable communication of $\ell$ elements of $\F_q$.

\begin{proof}

We check Privacy and Reliability.

\paragraph{{\it Privacy:}}
we have already observed in Proposition~\ref{protocol1:good} that Eve has no information on $\mathbf{h}\mathbf{y}^T$ (we drop the index $(i)$ to simplify notation), so that $z_1$ perfectly hides the secret. Now notice that if $\mathbf{y}$ can be decoded, then $z_2= s+\mathbf{h}\tilde{\mathbf{x}}^T= z_1-\mathbf{h}\tilde{\mathbf{e}}^T$; hence to conclude, it suffices to prove that Eve already knows whether $\mathbf{y}$ can be decoded or not, and that she knows $\tilde{\mathbf{e}}$ if $\mathbf{y}$ can be decoded. We prove this claim in the following lemma:

\begin{lemma}

Let $\mathbf{x}$ be a codeword sent by Bob to Alice, and let $\mathbf{y}= \mathbf{x}+\mathbf{e}$ be the received vector. Then Eve knows whether $\mathbf{y}$ can be decoded (i.e. $\mathbf{y}=\tilde{\mathbf{x}}+\tilde{\mathbf{e}}$ as above) or not; furthermore, if $\mathbf{y}$ can be decoded, then she knows $\tilde{\mathbf{e}}$.

\begin{proof}
By definition, $\tilde{\mathbf{e}}$ is a vector of minimum weight (and of weight at most $t/2$) such that $\mathbf{y}-\tilde{\mathbf{e}}$ belongs to $\C$; notice that the last condition is equivalent to require that $\mathbf{e}-\tilde{\mathbf{e}}$ belongs to $\C$. Now these requirements uniquely determine $\tilde{\mathbf{e}}$: indeed, if by contradiction $\mathbf{e}-\mathbf{e}'\in \C$ for another $\mathbf{e}'$, then $\mathbf{e}'-\tilde{\mathbf{e}}$ would belong to $\C$, a contradiction since $w_{\texttt{H}}(\mathbf{e}'-\tilde{\mathbf{e}})\leq t/2+t/2 < d_{\texttt{min}}(\C)$.

Hence $\tilde{\mathbf{e}}$ is uniquely determined by $\mathbf{e}$ and $\C$: Eve can thus compute it from the data in her possession. Notice that, in particular, she knows whether $\tilde{\mathbf{e}}$ exists or not, i.e. whether decoding of $\mathbf{y}$ is possible or not.
 
\end{proof}

\end{lemma}

\paragraph{{\it Reliability:}} we have two possible cases:

\begin{itemize}

\item if $|\mathcal{S}|\geq t/2$, then Bob is able to acquire the syndrome $\mathbf{H}\mathbf{y}^T$ of $\mathbf{y}$ via $t/2$-generalized broadcast (cf. Lemma~\ref{lemma:genbcast}); thus as remarked in Proposition~\ref{protocol1:good}, he can recover $\mathbf{y}$ and open the mask to get the secret.

\item if $|\mathcal{S}|<t/2$, then Bob knows that Alice has correctly decoded $\mathbf{y}$, since Eve introduced less than $d_{\texttt{min}}/2$ errors; thus $\tilde{\mathbf{x}}=\mathbf{x}$ so that $z_2 - \mathbf{h}\mathbf{x}^T= \left(s+\mathbf{h}\tilde{\mathbf{x}}^T\right) - \mathbf{h}\mathbf{x}^T= s$.

Notice that in this case Bob will have failed to decode the $t/2$-generalized broadcast but he will simply ignore the elements received in this way.

\end{itemize}

\end{proof}

\end{prop}

\begin{remark}\rm{
Notice that we could further improve the efficiency of this protocol
by requiring Alice to use $w$-generalized broadcast (instead of
regular one) to communicate the elements $z_1^{(i)}$ and $z_2^{(i)}$, where $w$ is the size of the
pseudo-basis; this, however, would not reduce the order of magnitude of the total cost.
}\end{remark}


\section{The Improved Protocol}\label{sec:protocol2}

The improved protocol simply implements the new techniques of sections~\ref{sec:basis3} and \ref{sec:masks}.

\begin{framed}

\begin{protocol}\label{protocol2}

The protocol allows Alice to communicate $\ell$ secret elements $s^{(1)},\cdots,s^{(\ell)}$ of $\F_q$ to Bob, where $q$ is an arbitrary integer with $q> n$. The protocol takes as input an MDS code $\C$ of parameters $[n,t+1,t+1]_q$ and a vector $\h$ of length $n$ as in Lemma~\ref{lemma:privcom}.

\begin{itemize}

\item[I.] Bob chooses $t+\ell+1$ uniformly random and independent codewords $\mathbf{x}^{(1)},\cdots,\mathbf{x}^{(t+\ell+1)}$ of $\C$ and sends them over the channels to Alice.

\item[II.] Alice receives the corrupted versions $\mathbf{y}^{(1)},\cdots,\mathbf{y}^{(t+\ell+1)}$, and she computes a pseudo-basis $\left\{\mathbf{y}^{(i)}:i\in I\right\}$ of the received values; she then proceeds with the following actions:

\begin{itemize}

\item[(i)] She uses Protocol~\ref{protocol:basis3} to communicate the pseudo-basis to Bob.

\item[(ii)] She then uses the remaining words to communicate to Bob the masked secrets and the data to retrieve them as in the first part of Protocol~\ref{protocol:masks}.

\end{itemize}

\item[III.] Upon receiving the pseudo-basis, Bob proceeds to compute the global support $\mathcal{S}$ of the error space; he can then obtain each secret $s^{(i)}$ as specified in the corresponding part of Protocol~\ref{protocol:masks}.

\end{itemize}
\end{protocol}
\end{framed}

Notice that privacy and reliability of the protocol follow from the previous discussions; we now analyze the complexity of the protocol:

\medskip

\noindent
\emph{Communication complexity:}
we can set $\ell:=1$.

\begin{itemize}

\item Step I requires transmitting $t+2$ words of $\F_q^n$ over the channels, thus requiring a total of $O\left(n^2 \right)$ symbols to be transmitted.

\item Step II-(i) requires transmitting $O\left( n^2 \right)$ elements of $\F_q$ as shown in Remark~\ref{rmk:basiscost}.

\item Finally, step II-(ii) requires using $t/2$-generalized broadcast to communicate $n$ symbols, and standard broadcast to communicate $2$ symbols, thus giving a total of $O(n)$ elements to be transmitted.

\end{itemize}

Hence since we can assume that $q = O(n)$, we get a total communication complexity of 

\[
O\left( n^2\log n \right)
\]

bits to be transmitted to communicate a single-bit secret.

\medskip

\noindent
\emph{Transfer rate:} optimal rate is achieved for $\ell=\Omega\left( n
\right)$.

\begin{itemize}

\item Step I requires transmitting $t+\ell+1 = \ell+O(n)$ codewords, for a total of $n\ell+O(n^2)$ symbols.

\item Step II-(i) remains unchanged from the single-bit case, and thus requires transmitting $O\left( n^2 \right)$ symbols.

\item Finally, step II-(ii) uses $t/2$-generalized broadcast to communicate $\ell t$ elements of $\F_q$ and standard broadcast to communicate $2\ell$ elements of $\F_q$, so that the overall cost is equal to $4 n\ell$ symbols to be transmitted.

\end{itemize}

To sum up, the overall transmission rate is equal to
\[
\frac{5n\ell+ O\left( n^2 \right)}{\ell} = 5n + O\left(n^2/ \ell\right)
\]

Furthermore, by using Reed-Solomon codes (instead of arbitrary MDS ones), we then have that Protocol~\ref{protocol2} has computational cost polynomial in $n$ for both Alice and Bob.


\section{Generalization to Linear Combinations of Errors and Eavesdropped Data}
\label{sec:linear}

We present in this section a more general scenario, where Alice and Bob can communicate words to each other in the presence of a more powerful adversary Eve, who can eavesdrop linear combinations of the transmitted symbols and inject linear combinations of errors from a set of her choice.
We show that Protocol~\ref{protocol1} (our vanilla protocol) can be generalized in order to provide security in this more complex scenario; this shows how our protocol carries greater potential for more general settings compared to previous work that relied on more cumbersome communication techniques.

\smallskip


The generalization of the communication model is defined as follows:

\begin{definition}[\textbf{Generalized Model}]\label{defi:generalized_adv}

Let $m, n,t$ be integers, and $\F_q$ be a finite field. Alice and Bob can communicate to each other $n$-tuples $\x=[\x_1,\ldots, \x_n]$ where each $\x_i$ is a column vector in~$\F_q^m$.

The adversary Eve is computationally unbounded, and selects at the beginning of the protocol $t$ ``eavesdropping vectors" $\lam^{(1)},\ldots,\lam^{(t)}\in \F_q^n$ and $t$ ``tampering vectors" $\fatmu^{(1)},\ldots,\fatmu^{(t)}\in \F_q^n$.

\noindent
Whenever $\x=[\x_1,\ldots, \x_n]$ is transmitted (either from Alice to Bob or from Bob to Alice), the following happens:

\begin{itemize}

\item
Eve learns the value of

\[
\lam^{(i)}\x^T= \lambda_1^{(i)}\x_1+ \cdots +\lambda_n^{(i)}\x_n
\]

for every $i=1,\ldots, t$.

\item
Eve selects $t$ columns vectors $\d^{(1)},\cdots,\d^{(t)} \in \F_q^m$; the intended receiver gets the message $\y=\x+ \e$, where 

\[
\e
= \sum_{i=1,\ldots, t} \d^{(i)} \otimes \fatmu^{(i)}
= \sum_{i=1,\ldots, t} \left[ \d^{(i)} \mu_1^{(i)},\cdots, \d^{(i)} \mu_n^{(i)} \right] 
\, .
\]

\end{itemize}

\end{definition}

Notice that PSMT can be seen as a more restrictive version of this model, where Eve is forced to choose vectors $\lam$ and $\fatmu$ of weight $1$, and where moreover $\lam^{(i)}= \fatmu^{(i)}$ for every $i$.

Though the adversary of Definition~\ref{defi:generalized_adv} is way more powerful than the one in classical PSMT, we show that Protocol~\ref{protocol1} can be adapted to provide security in this scenario as well.
More precisely, the only relevant modification we have to perform is the following: instead of classical codes, which can correct from errors with bounded Hamming weight, we use \emph{rank} codes which can correct from errors with bounded rank; notice that such are the errors introduced by the adversary in Definition~\ref{defi:generalized_adv}.

\smallskip

As a final remark, we stress the fact that this more complex scenario
is not a purely gratuitous generalization but
has an application to Secure Network Coding.

Indeed, in  Network Coding, one or several transmitters are connected to one or several receivers by a network, i.e. a directed multigraph with  source nodes and destination nodes; ``network coding" means that each node performs $\F_q$-linear operations on the symbols received via the incoming edges, and sends the results through the outbound edges.
Each sender can feed input to its corresponding source node, and each receiver can read the output of the corresponding destination node.
In the ``Secure Network Coding'' scenario (e.g. \cite{SilvaKIEEE}), we have
typically a single transmitter and
an adversary that controls $t$ edges, meaning he can read the symbols transmitted over these edges and replace them by symbols of his choice.

In case there is a single receiver, we can identify the sender with Alice and the receiver with Bob; it is then readily seen that the communication between Alice and Bob is affected by the adversary precisely as in Definition~\ref{defi:generalized_adv}.
This means that our protocol can be used to provide security for a Network Coding scenario with a single receiver, as long as communication is also possible from the receiver to the sender. Though
Network Coding was originally introduced in a multicast scenario~\cite{Netcode1,Netcode2}, it has since been proved useful in single sender - single receiver scenarios as well \cite{Unicast1,Unicast2}.

\smallskip

Until very recently, existing work on Secure Network Coding assumes that information can only flow from sender to receiver. Notably, the  work of Silva and Kschischang~\cite{SilvaKIEEE} presents a one-round protocol that is secure as long as $t<n/3$;
recently, the present authors introduced a protocol~\cite{SpiniZemor} that achieves security for any $t<n/2$, in a multiple receiver context, 
by allowing communication from receiver to sender as well.
This protocol uses three rounds of communication.
We sketch below how a two-round generalization of Protocol~\ref{protocol1} is also secure for any $t<n/2$. This can be directly applied to Secure Network Coding in a unicast (single transmitter -- single receiver) scenario.

We make the final remark that Jaggi et al.~\cite{SecureCoding} studied a similar case, where the adversary controls vertices instead of edges of the network; their protocol lets the adversary inject up to $n/2$ errors, but with a weaker notion of security in that it must drop the privacy requirement and the reconstruction process admits a positive error probability.



\subsection{Communication with Rank-Metric Codes}\label{sec:rank}

We show in this section how the machinery of Protocol~\ref{protocol1} can be adapted to another type of code, defined under the \emph{rank metric}~\cite{Delsarte},~\cite{Gabidulin}. These have been extensively used in Secure Network Coding~\cite{SilvaKWiretap},~\cite{SilvaKIEEE}.

\begin{definition}[Rank-Metric Code]\label{defi:MRD}

Given the $\F_q$-linear space $\F_q^{m\times n}$ ($m$-by-$n$ matrices over $\F_q$), we can define the \emph{rank distance} between its elements by letting $d_R(\x,\y):=\text{rank}(\y-\x)$.

A \emph{rank-metric} code $\C$ is a non-empty subset of $\F_q^{m\times n}$ with induced rank distance; by identifying the field $\F_{q^m}$ with $\F_q^m$, we can view $\C$ as a code over $\F_{q^m}^n$, and require it to be \emph{linear} over $\F_{q^m}$; we can hence speak of \emph{block length} and \emph{dimension} of such a code (as in the Hamming case) and of \emph{minimum (rank) distance}, and combine these parameters into the triplet $[n,k,d]_{q^m}$.

\end{definition}

The equivalent concept of MDS in this setting is called \emph{Maximum Rank-Distance}: a rank-metric code is Maximum Rank-Distance (or \emph{MRD} for short) if $k+d=n+1$; an MRD code of arbitrary dimension $k$ and length $n$ exists if and only if $m\geq n$~\cite{Delsarte}. More precisely, for any $q$ and any $m,n,k$ with $m\geq n$, we can construct a \emph{Gabidulin code}~\cite{Gabidulin} of parameters $[n,k,n-k+1]_{q^m}$.

As in the Hamming case, we can express a rank-metric code~$\C$ in term of a \emph{parity-check} matrix $\H$, i.e.

\[
\C= \left\{ \x\in \F_{q^m}^n: \, \H \x^T= \mathbf{0} \right\}.
\]

Furthermore, we can define the associated \emph{syndrome} map $\sigma:\w  \mapsto \H \w^T$; clearly, we have that $\C = \ker (\sigma)$.

%
%
%
%

As a first building block, we construct a rank-metric equivalent of Lemma~\ref{lemma:privcom}: we show how to define a rank-metric code $\C$ and a vector $\h$ such that the adversary has no information of the value of $\h\x^T$ if a random $\x\in\C$ is transmitted between Alice and Bob. Here the key point is that the vector $\h$ has entries in $\F_{q^m}$ rather than in $\F_q$.

\begin{lemma}\label{lemma:rankprivcom}

For any $q$, $n$, $m\geq n+1$ and $t<n$ there exists an MRD code $\C$ of parameters $[n,t+1,n-t]_{q^m}$ and a vector $\h\in \F_{q^m}^n$ such that given a random codeword $\x\in\C$ and $t$ arbitrary vectors $\lam^{(1)},\cdots, \lam^{(t)}  \in \F_q^n$, the value of $\h\x^T$ is completely undetermined even if $\lam^{(1)} \x^T, \cdots, \lam^{(t)} \x^T$ are known.

\begin{proof}

We construct $\C$ and $\h$ by adapting the blueprint of Lemma~\ref{lemma:privcom} to the rank-metric setting: hence we first let $\C'$ be an MRD code of parameters $[n+1,t+1,n-t+1]$, and let

\[
\C:= \left\{ \x\in \F_{q^m}^n:\exists \, x_{n+1}\in \F_{q^m} \text{ with } (\x,x_{n+1})\in \C' \right\}
\]

We then define $\h$ by selecting a parity-check matrix $\H'$ of $\C'$, and choosing a row thereof of the form

\[
(\h, \alpha)\in \F_{q^m}^{n+1} \text{ with } \h\in \F_{q^m}^n, \, 0\neq \alpha\in \F_{q^m}.
\]

Now consider the following matrix:

\[
\mathbf{M}:=
\left[\begin{array}{ccc}
&\H'& \\ \hline
\lam^{(1)} &|& 0 \\
\cdots &|& 0 \\
\lam^{(t)} &|& 0 \\ \hline
\mathbf{0} &|& -\alpha
\end{array}\right]
\cdot \left[ \begin{array}{c}
\x_1 \\
\vdots \\
\x_n \\
x_{n+1}
\end{array} \right]
\]

clearly, if $M$ is non-singular then the claim holds (we assume as a worst-case scenario that the vectors $(\lam^{(i)}:i=1,\cdots, t)$ are linearly independent). Now by properties of MRD codes, the matrix
$\begin{bmatrix} \H \\ \B \end{bmatrix}$ 
is non-singular for any full-rank matrix $\B\in \F_q^{t\times n}$~\cite{SilvaKWiretap}; hence in particular $\mathbf{M}$ is of full-rank, so that the claim holds.

\end{proof} 
\end{lemma}

In a symmetric fashion, we now present a rank-metric version of the broadcast protocol, hence allowing for reliable communication of messages, although with no guarantee of privacy.

\begin{lemma}\label{lemma:rankbcast}

Given any $q$ and $n\leq m$, let $\C_{\texttt{BCAST}}$ be an MRD code of parameters $[n,1,n]_{q^m}$. Then if an arbitrary $\x\in \C_{\texttt{BCAST}}$ is transmitted between the players, the receiver can always recover $\x$ from the received message $\y$ by computing the closest codeword to $\y$.

\begin{proof}

By Definition~\ref{defi:generalized_adv}, the receiver obtains a message $\y$ with $\y=\x+\e$, where the error $\e$ introduced by Eve is of rank at most $t$. Hence the original codeword $\x$ can be recovered since $\C_{\texttt{BCAST}}$ has rank-distance $n\geq 2t+1$.

\end{proof}
\end{lemma}

Furthermore, we see that the machinery of the pseudo-basis can be adapted to the rank-metric case; we begin with the equivalent of Lemma~\ref{lemma:errweight}:

\begin{lemma}\label{lemma:inj}

Let $\C$ be a rank-metric code of parameters $[n,k,d]_{q^m}$; let $\H$ be a parity-check matrix of $\C$, and let $W\leq \F_{q^m}^n$ be an $\F_{q^m}$-linear subspace with the property that each $\mathbf{w}\in W$ is of rank at most $d-1$ over $\F_q$. Then the syndrome map $\sigma:\w \mapsto \H  \w^T$ is injective on $W$.

%
%
%

\end{lemma}

As in the Hamming case, given a rank-metric code $\C$ of length $n$ over $\F_q$ and a set $\mathcal{Y}$ of vectors in $\F_{q^m}^n$, we call a pseudo-basis of $\mathcal{Y}$ a subset $\mathcal{W}\subseteq \mathcal{Y}$ such that $\sigma(\mathcal{W})$ is a basis of $\langle \sigma(\mathcal{Y})\rangle$, where $\sigma$ denotes a syndrome map of $\C$.

Again, we have that since the codomain of $\sigma$ is equal to $\F_{q^m}^{n-k}$, a pseudo-basis has cardinality at most $n-k$; furthermore, a pseudo-basis can be computed in time polynomial in $n$.

We now show that if the set $\mathcal{Y}$ consists of corrupted codewords, affected by errors introduce by the adversary, then a pseudo-basis corresponds to a basis of the error space in this setting as well; 
the proof follows the same steps as that of Proposition~\ref{proposition:pseudo-basis}:

\begin{prop}\label{prop:rank_pseudo-basis}

Let $\C$ be a linear rank-metric code of parameters $[n,k,d]_{q^m}$, and let $\mathcal{X}$, $\mathcal{E}$, $\mathcal{Y}$ be three subsets:

\[
\begin{array}{rll}
\mathcal{X} := & \left\{ \mathbf{x}^{(1)},\cdots,\mathbf{x}^{(r)}  \right\} & \subseteq \C, \\
\mathcal{E} := & \left\{ \mathbf{e}^{(1)},\cdots,\mathbf{e}^{(r)}
                 \right\} & \subseteq \F_q^n\quad \text{such that } \text{rank}(\e)\leq d-1 \text{ for all } \e \in \langle \mathcal{E}\rangle_{\F_{q^m}}, \\
\mathcal{Y} := & \left\{ \mathbf{y}^{(1)},\cdots,\mathbf{y}^{(r)}
                 \right\} & \subseteq \F_q^n\quad \text{with } \y^{(j)}=\x^{(j)}+\e^{(j)} \text{ for every } j
\end{array}
\]

Then, given knowledge of $\X$ and a pseudo-basis of $\Y$, we can compute $\mathbf{e}^{(j)}$ from its syndrome $\sigma(\mathbf{e}^{(j)})$, for any $1\leq j\leq r$.

\end{prop}

We show in the following section how to implement the techniques we presented.

\subsection{The Protocol for The Rank-Metric Case}\label{sec:rank-metric_protocol}

We define in this section the protocol for private and reliable communication in the setting of Definition~\ref{defi:generalized_adv}.

\begin{framed}

\begin{protocol}\label{protocol3}

We assume that Alice and Bob can communicate vectors to each other in the presence of an adversary Eve as in Definition~\ref{defi:generalized_adv}; we assume that $t<n/2$ and that $m>n$.

The protocol allows Alice to communicate $\ell$ secret elements $s^{(1)},\cdots,s^{(\ell)}$ of $\F_{q^m}$ to Bob, and it takes as input a pair $(\C,\h)$ as in Lemma~\ref{lemma:rankprivcom} for private communication, and a rank-metric code $\C_{\texttt{BCAST}}$ for reliable communication as in Lemma~\ref{lemma:rankbcast}.

\begin{itemize}

\item[I.] Bob chooses $t+\ell$ uniformly random and independent codewords $\mathbf{x}^{(1)},\cdots,\mathbf{x}^{(t+\ell)}$ of $\C$ and communicates them to Alice.

\item[II.] Alice receives the corrupted versions 
$\mathbf{y}^{(1)}=\x^{(1)}+\e^{(1)},\cdots,\mathbf{y}^{(t+\ell)}=\x^{(t+\ell)}+\e^{(t+\ell)}$; she then proceeds with the following actions:

\begin{itemize}

\item[(i)] She computes a pseudo-basis $\left\{\mathbf{y}^{(i)}:i\in I\right\}$ for $I\subset \{1,\cdots,t+\ell\}$ of the received values, and uses $\C_{\texttt{BCAST}}$ to reliably communicate to Bob $\left(i,\mathbf{y}^{(i)}:i\in I\right)$.

\item[(ii)] She then considers the first $\ell$ words that do \emph{not} belong to the pseudo-basis; to ease the notation, we will re-name them $\mathbf{y}^{(1)},\cdots,\mathbf{y}^{(\ell)}$. For each secret $s^{(j)}$ to be communicated she broadcasts to Bob the following two elements:

\begin{itemize}

\item[-] $\mathbf{H}\left( \mathbf{y}^{(j)} \right)^T$, the syndrome of $\y^{(j)}$;

\item[-] $s^{(j)}+\mathbf{h}\left( \mathbf{y}^{(j)} \right)^T$.

\end{itemize} 

\end{itemize}

\item[III.] 

Proposition~\ref{prop:rank_pseudo-basis} guarantees that for any $j$, $1\leq j\leq \ell$, Bob can compute the error vector $\e^{(j)}$ and hence reconstruct $\y^{(j)}=\x^{(j)}+\e^{(j)}$ from his knowledge of $\x^{(j)}$. He can therefore open the mask $\mathbf{h}\left( \mathbf{y}^{(j)} \right)^T$ and obtain the secret $s^{(j)}$. 

\end{itemize}

\end{protocol}

\end{framed}

The security of Protocol~\ref{protocol3} can be proved by adapting the proof of Proposition~\ref{protocol1:good} to the rank-metric case. Furthermore, Protocol~\ref{protocol3} has polynomial cost in $n,q$ and $m$ for both computation and communication.

%


\section{Concluding Remarks}\label{sec:conclusion}

We have presented a two-round PSMT protocol that has polynomial computational cost for both sender and receiver, and that achieves transmission rate linear in $n$ and communication complexity in $O\left( n^2 \log n \right)$; we believe that our protocol is conceptually simpler compared to previous work and fully harnesses the properties of the pseudo-basis.

As proved in \cite{Srinathan}, the transfer rate is asymptotically optimal; furthermore, our protocol has a low multiplicative constant of $5$. We moreover show that our vanilla protocol can be adapted to more general scenario, in the presence of a more powerful adversary.

It remains open whether the $O\left(n^2\log n\right)$ communication complexity is optimal or not; the only known lower bound on this parameter is still $O(n)$, as the one for transfer rate \cite{Srinathan}. It seems to us that a communication complexity lower than $O\left( n^2 \right)$ is unlikely to be achievable, at least not without a completely different approach to the problem.


\section{Acknowledgments}

The authors would like to thank Serge Fehr, Ronald Cramer and Muriel Médard for their useful comments and suggestions.



\end{document}